\PassOptionsToPackage{table}{xcolor}
\documentclass[sigconf]{acmart}
\usepackage{colortbl} % row colors
\usepackage[capitalise]{cleveref}
\usepackage{tcolorbox}

\usepackage{framed}
\usepackage{changepage}

\newcommand{\interviewquote}[2]{
 \def\FrameCommand{%
    \hspace{0pt}%
    {\color{cyan} \vrule width 2pt}% <-- Change color here.
    \colorbox{white}
  }%
  \MakeFramed{\advance\hsize-\width\FrameRestore}%
  \noindent% disable indenting first paragraph
  \begin{adjustwidth}{}{1pt}
  {\small``\textit{#1}'' - {#2}}\end{adjustwidth}\endMakeFramed%
}

\AtBeginDocument{%
  }

\copyrightyear{2026}
\acmYear{2026}
\setcopyright{cc}
\setcctype{by-nc-nd}
\acmConference[ICSE-SEET '26]{2026 IEEE/ACM 48th International Conference on Software Engineering}{April 12--18, 2026}{Rio de Janeiro, Brazil}
\acmBooktitle{2026 IEEE/ACM 48th International Conference on Software Engineering (ICSE-SEET ‘26), April 12--18, 2026, Rio de Janeiro, Brazil}
\acmPrice{}
\acmDOI{10.1145/3786580.3786980}
\acmISBN{979-8-4007-2423-7/2026/04}

\begin{document}

\title{Integrating Mental Health, Well-Being, and Sustainability into Software Engineering Education}

\author{Isabella Gra{\ss}l}
\email{isabella.grassl@tu-darmstadt.de}
\orcid{0000-0001-5522-7737}
\affiliation{%
  \institution{Technical University of Darmstadt}
  \city{Darmstadt}
  \country{Germany}}

\author{Birgit Penzenstadler}
\email{birgitp@chalmers.se}
\orcid{0000-0002-5771-0455}
\affiliation{%
  \institution{Chalmers University of Technology}
  \city{Gothenburg}
  \country{Sweden}}

\begin{abstract}
Mental health and well-being are major concerns in higher education and professional fields such as software engineering, yet are often overlooked in curricula.
This paper describes our approach to include mental health, well-being, and sustainability in software engineering education in two ways: (1) well-being-focused software projects that ask students to design technical solutions or research addressing mental health and sustainability or societal challenges, and (2) brief classroom interventions such as short reflective discussions and team-building activities. 
We argue that this combination can help students see software engineering more broadly while creating healthier learning environments. Our analysis of reflections from 60 students found several positive outcomes:  
students gained a more human-centred perspective, had more team discussions about mental health, and began to see well-being as inspiration for using software to benefit society and individuals rather than merely as a technical or business tool. 
By combining technical skills with awareness of well-being, we argue that software engineering education can prepare future developers to be both skilled programmers and responsible professionals who care about human well-being.
\end{abstract}




\maketitle

\section{Introduction}
%\igtodo{@Birgit: What’s your opinion on starting with a quote from the students? I like ‘unconventional’ openings, and I guess it would be appropriate here, but would love to hear your opinion.}\bptodo{Ja! super!} 
%\ig{sag auch gern, falls du eine andere quote passender findest!}

\interviewquote{I felt like a coding machine in all those pure technical seminars and projects, and with this [course], I realized that there is more to SE and that it is okay to feel overwhelmed and down and just talk to the others.}{ID57}

This reflection\footnote{\emph{SE} means software engineering. \emph{ID} followed by a number represents a student from our courses.} illustrates how students often experience software engineering education as narrowly technical, with little space to acknowledge their own emotions or broader perspectives.  
Such experiences reflect the field’s dominant instrumental orientation: an emphasis on tools and technical mastery as well as commercial value at the expense of human concerns and values~\cite{groeneveld2019}. A humanistic orientation, by contrast, emphasises empathy, responsibility, and care for learners as whole persons as well as putting the user at the very centre~\cite{sarker2019}. Balancing these two orientations is essential~\cite{lee2004} to prepare software engineers as reflective professionals aware of the impact of their work and their own well-being~\cite{groeneveld2022b,takaoka2024}.

At the same time, student well-being and mental health have become urgent issues in higher education~\cite{korsten2021,lipson2022}, resulting in a \emph{mental health crisis}~\cite{stokoe2024}. Gen Z and Alpha\footnote{Generation Z (late 1990s–2010) are today’s undergraduates, while Generation Alpha (2010 onwards) are the upcoming students entering higher education.} report higher levels of stress, anxiety, and depression than previous cohorts~\cite{sanchez-pena2025}, shaped by competition and pressure, social media, and uncertainty about the future. 
While these challenges affect students across disciplines, they also intersect with the \emph{culture of engineering}~\cite{asghar2024}, which often discourages open discussion of mental health and frames struggling as a personal weakness~\cite{mcalister2023,asghar2024}. These are linked to stereotypically masculine norms shaped by society that can further inhibit help-seeking behaviour~\cite{wright2021}, which might be especially relevant for under-represented groups~\cite{pigart2025,vahidi2024,danowitz2018}.

These challenges are not only personal but become professional. Software engineers design systems that shape everyday life and need to be aware of human needs~\cite{walther2017}. If education frames software engineering as a purely technical pursuit, it risks producing graduates who lack the reflective capacity to consider well-being, both their own and that of the societies their systems affect~\cite{moreira2024,lee2004,takaoka2024}.

This gap is concerning, but also offers an opportunity to rethink how we design courses. By embedding sustainability, well-being, and mental health into the curriculum, we can support students’ personal development~\cite{li2025b}, and broaden their view of software engineering’s societal role~\cite{moreira2024}.

This experience report presents our approach to integrating mental health, well-being, and sustainability into undergraduate and graduate software engineering courses. We combine socially relevant projects with low-key classroom practices such as reflections and team-building activities. 
Our intention is twofold: to foster students’ personal well-being, and to foster a generation of software engineers who see themselves as agents of change, equipped with both technical skills and human values.

\begin{tcolorbox}[
  colback=cyan!3,    % light grey background for the box
  colframe=cyan,   % border color
  colbacktitle=cyan, % darker grey background for the title
  coltitle=white,     % title text color
  title=Contribution, % the title text
  fonttitle=\bfseries % bold title
]
Our findings illustrate how students progressed in expanding their stereotypical view of software engineering, cultivating empathy for users, teammates, and themselves, and recognising their role in shaping society. We provide guidance for educators to develop technical competence and social responsibility in future software engineers.
\end{tcolorbox}

\section{Background and Related Work}

This section reviews how mental health, well-being, and sustainability are understood and addressed in software engineering, both in professional and educational contexts, to situate our contribution.

\subsection{Mental Health, Well-Being, Sustainability}
We begin by clarifying how we understand key terms and why they matter for software engineering and education.

\subsubsection{Definitions}
Sustainability is not just environmental; it includes social and human sustainability. Mental health and well-being are explicitly part of the UN Sustainable Development Goals (\emph{UN SDG}, \emph{SDG 3: Good Health and Well-being})\footnote{\url{https://sdgs.un.org/goals}}. Further issues, such as social justice and climate change, related to the \emph{UN SDGs} are affected and shaped by software engineers. Thus, mental health should be important for (future) software~\cite{takaoka2024}.
The World Health Organization (WHO) defines mental health as “a state of mental well-being that enables people to cope with the stresses of life, realise their abilities, learn well and work well, and contribute to their community,” and notes that this is more than the absence of mental disorders~\cite{worldhealthorganization2022}. 
This perspective includes both negative aspects, such as distress, and positive aspects, such as resilience. 

Well-being\footnote{Well-being can be seen both as a moderator (impacting work satisfaction and performance) and as a mediator (affecting hedonic well-being)~\cite{godliauskas2025}.} forms a foundation for effective learning and professional practice: students cannot contribute fully to software products or society if their own well-being is neglected~\cite{takaoka2024}. In education, integrating well-being and sustainability represents an emerging approach to mindfulness education~\cite{li2025b}. By attending to students’ emotions and social skills, courses can enhance emotional regulation, teamwork, collaborative abilities, and self-confidence, skills that are particularly important in software engineering.

\subsubsection{The Profession of Software Engineering}
Software developers consistently report higher anxiety than professionals in other sectors~\cite{nayak2014}. Work environments are often characterised by high demand, time pressure, and social isolation~\cite{graziotin2019,aires2024}. Ford et al. were among the first to explore emotional states and negative experiences in software development~\cite{ford2015exploring}, followed by Graziotin et al. on developer (un)happiness~\cite{graziotin2018happens}, and research on burnout~\cite{tulili2023burnout,raman2020stress}.

Cultural and stereotypical perceptions and assumptions of software engineering, such as a strong focus on technical expertise, rationality, and masculine norms, mirror broader engineering culture~\cite{vera2021}. This is reflected in how communication channels shape and challenge a participatory culture in software development~\cite{storey2016social}, confirmed by findings by Cheriyan et al. on the detection and reduction of offensive language in software engineering~\cite{cheriyan2021towards}.
Further studies highlight factors for mental health~\cite{wong2023mental} and wellbeing~\cite{montes2025wellbeing} and conclude that an integration of these concepts into software engineering education would benefit the profession~\cite{takaoka2024}.

\subsection{Educational Settings}
Building on the professional context, we turn to educational settings where the next generation of professionals is trained.

\subsubsection{Approaches in Engineering and STEM}
Research on student well-being in higher education has often focused on engineering and STEM broadly, offering useful insights but rarely addressing software engineering directly.  
Most existing studies on engineering students’ well-being have been conducted in North America, with fewer in Europe or the Global South~\cite{asghar2024}. Research has highlighted that engineering students face significant academic stress, challenges in balancing responsibilities, and pressures from a competitive culture. Educators often recognise these stressors but are less aware of interpersonal or identity-related stressors, as these do not fall within their areas of responsibility~\cite{sanders2025,sanders2024}.

The \emph{engineering culture}~\cite{asghar2024} itself is a barrier to addressing mental health as it tends to reward toughness, rationality, and working through pain, which are traditionally masculine traits that discourage open emotional expression or vulnerability~\cite{sanders2024,godfrey2010}. 

Several interventions have been conducted to support mental health and well-being in engineering education, ranging from short mindfulness activities~\cite{huerta2018,joshi2016,miller2020,aree2020}, academic study skills support~\cite{andrews2020}, and stress management workshops~\cite{terrell2025}, to embedding well-being into seminars through storytelling or open discussion of burnout~\cite{grozic2025}. 
The behaviour of educators also plays an important role: showing empathy, using supportive language, and modelling healthy coping strategies can promote trust~\cite{harper2024,busch2024}. 

These factors are particularly important for underrepresented groups such as women, Black, Latin, and Indigenous students, who report higher stress and anxiety~\cite{pigart2025,vahidi2024,danowitz2018}. 
Early-year students may also require specific support to address loneliness~\cite{abidin2025}. However, barriers to help-seeking at university remain widespread in engineering contexts~\cite{wright2023}, with students desiring a structured, institutional support~\cite{folkman2004,navarro2025,jensen2023}.

\subsubsection{Software Engineering Courses}
Students’ well-being in software engineering courses has received far less attention compared to other engineering disciplines. 
This is a relevant gap, given that students in computing face the highest risk of mental health disorders among all engineering disciplines~\cite{danowitz2018}. 

Stressors specific to software education include challenges in debugging~\cite{chandrasekaran2025} and high rates of imposter syndrome~\cite{rosenstein2020}. Students frequently report anxiety, burnout, and lack of self-awareness of their stress levels~\cite{syed-abdullah2006}. Teamwork in software courses can provide opportunities to experience positive stress and resilience~\cite{grassl2023k}, while requiring psychological safety to ensure belonging and reduce dropout risks~\cite{cooke2023,andrews2020}. These challenges are especially pronounced for first-year undergraduates, who often face transition-related stress, loneliness, and limited coping strategies~\cite{sanchez-pena2025,asghar2024}. 

Despite these risks, software engineering education has largely overlooked student well-being. Thus, we explore how to pragmatically integrate mental health, well-being, and sustainability into undergraduate and graduate software engineering courses.

\section{Methods}
We describe the educational setting, data collection, and analysis procedures used to explore students’ experiences of well-being, mental health, and sustainability in software engineering courses.

\subsection{Course Context}
The courses took place at the Technical University of Darmstadt, Department of Computer Science, in the summer term of 2025. 
We collected written reflections from students enrolled in three elective courses within the module \emph{Software and Hardware} (also referred to as \emph{Software Systems and Formal Methods}). 
Two courses were research-oriented seminars, and one was a practical software engineering lab. 
Although situated in computer science, these elective courses are open to students from other programs such as Cognitive Science or Business Informatics. 
At the university, most electives are available to both undergraduate and graduate students. This mixture can influence the dynamics of the courses, as graduate students may bring more advanced research or project experience. In our view, this diversity made the reflections on well-being and sustainability richer and authentic.

\subsubsection{Rationale for Course Selection}
We deliberately selected these three courses to cover two different but complementary formats (1) \emph{Research seminars} (2 courses) with focus on academic reading, conducting research, writing a paper, and presenting in a team, and (2) \emph{Practical lab} (1 course) with focus on user-centred design, the creation of a working prototype, documenting the process in a report and presenting it in a team. 

By including both formats, we aimed to assess how integrating well-being, mental health, and sustainability goals would be beneficial in both theoretical and hands-on contexts. 
The learning objectives and sustainability aims were aligned across both formats, but the outcomes differed. 
This combination also allowed us to assess whether the integration of well-being and sustainability worked consistently across settings or required format-specific adaptations.

\subsubsection{Shared Pedagogical Elements}
Across the courses, we integrated several elements that reflected our shared learning goals of connecting software engineering with well-being, mental health, and the UN SDGs. The activities were deliberately designed to be carried out in both research seminars and the practical lab, even though the course formats differ.

\paragraph{Topics: Connection to UN SDGs}
The topic selection is a key element in our courses. In both the seminar and the lab, students should find something that interests \emph{them}. They had creative freedom; the only restriction was that it should be human-centred and related to sustainability. To make this clearer, we provided an introduction to the \emph{UN SDGs} as a guiding framework\footnote{\url{https://sdgs.un.org/goals}}. Students were encouraged to relate their projects or research questions to at least one of the goals, such as SDG 3 (Good Health and Well-being) or SDG 13 (Climate Action). The aim was not strict alignment, but rather to help students see the broader societal relevance of their work and to reflect on how software can contribute to sustainable development.
We provided students with examples from the real world, e.g., specific research papers or applications for social good, such as cleaning up the ocean or local wildlife.

\paragraph{Empowering Real-world Examples} %wording..
At the beginning of every in-person session, we started with a short positive example from software development or society. There were two main reasons for this. First, we wanted to create a hopeful, constructive atmosphere because the world often feels overwhelming, and we wanted to position them as agents and problem-solvers. Second, we wanted to show what is currently happening in research or society, and how this connects to software, inclusion, and sustainability. 
Examples included a BBC video about using glitter to track endangered water voles,\footnote{\url{https://www.bbc.com/news/videos/c9q04x2wdlzo}} a Guardian article about Salt Lake City adopting Pride flags in response to political restrictions\footnote{\url{https://www.theguardian.com/us-news/2025/may/07/salt-lake-city-boise-pride-flags}}, queer successful people in software engineering (also how to define success), information about local mental health events at the university. 
These examples were intentionally unconventional to emphasise that unusual or \emph{outside-the-box} ideas can be valuable in both research and practice, and that students should not be afraid to share ideas, even if they seem unlikely at first.

\paragraph{Team Contract}
Teamwork was a central aspect in all courses. Every team was required to create a team contract together, in person. We provided a template with initial prompts such as: \emph{What are our goals? How do we ensure accountability? How do we want to communicate? How do we handle conflict? What solutions and consequences do we agree on?}
Students could adapt the template, but had to discuss these questions and document their agreements. This exercise not only clarified expectations, but also promoted healthy team dynamics, reducing stress and misunderstandings. By planning how to communicate and resolve conflicts, students learned strategies that support mental well-being and sustainable collaboration throughout their projects.

\paragraph{Teamwork and Communication}
All courses included explicit sessions on teamwork, communication, and conflict resolution. We emphasised that successful software development, whether in research or industry, requires collaboration, empathy, and self-regulation. Students were encouraged to reflect on their own working styles and share experiences from studies or extracurricular activities such as sports.
We also integrated evidence-based resources to support inclusive and sustainable teamwork. For example, students listened to a podcast by Andrew Begel on neurodivergence and communication in software teams.\footnote{\url{https://csedpodcast.org/blog/s4e10_tpb_team_communication_neurodivergence/}}

\emph{Reflections.}
Reflection was emphasised in all courses. Teams wrote weekly reflections that went beyond reporting progress and explicitly invited students to write about their reasoning and feelings during the week, no matter whether they felt confident, anxious, motivated, or frustrated, and why. We gave guidelines such as: be honest (failures are as important as successes), write with depth (focus on one or two aspects in detail), and be professional (think about the audience). The reflections were intended as a space to connect technical and non-technical experiences, including teamwork and emotions.

\paragraph{Feedback-Loop}
The instructor also reflected openly in class. At the beginning of each session, she presented a short feedback loop slide with questions like: \emph{How is your progress? What challenges did you face? What topics do you want to discuss?} Alongside this, she shared her own learnings so far, for example, realising that some students needed more scaffolding and responding by adding templates and structured examples. This transparency underlined that learning was a shared process and that there are no ‘dumb questions’, only unasked ones.

\subsubsection{Research Seminars}
We provide the full list of topics, literature, and all seminar and lab materials in our replication package.

%\igtodo{paper references, description, concrete schedule (abstract version, e.g. table with W1, W2, ...)}
%check space first!

\paragraph{Seminar on \textbf{Diversity in SE} (4 ECTS)}
Students worked in pairs to investigate topics around diversity in software engineering and to write a scientific paper. 
The seminar emphasised critical engagement with research findings, identifying research gaps, and following academic standards for writing and reviewing. 
Students also practised presenting their work and providing constructive peer feedback. 
Starting topics included inclusive programming education, diversity in software teams, bias and fairness in AI and software systems, neurodiversity and cognitive inclusion, queerness and sexuality in software development, mental health and well-being in software engineering, and other emerging topics published during the summer term.

The seminar began with a topic assignment, team formation, and a kick-off session. Students attended workshops on \emph{How to Research and Write}, writing practice sessions, elevator pitches, presentation skills, and teamwork. They also had three opportunities for individual meetings with the instructor. The final outcome was a scientific paper on a topic of their choice, encompassing the full research process: investigation, writing, presenting, and reviewing.

\paragraph{Seminar on AI in SE (2 groups, 4 ECTS)}
Two student groups from \emph{AI in SE} were supervised by the instructor of the main seminar. This seminar was more technically oriented than \emph{Diversity in SE}, but students received the same guidance on research and paper writing. Sustainability and well-being topics were also included to test whether these elements could be applied to a more technical research context.

\subsubsection{Practical Lab on Society and SE (6 ECTS)}
Students worked in teams of 4–6 to design software solutions for local or global societal challenges. The lab followed a Design Thinking approach, focusing on teamwork, user needs, and prototyping rather than technical implementation. The aim was to create No-Code or Low-Code prototypes suitable for participants with varying technical skills, resulting in a \emph{minimal lovable product}.

Activities included topic introduction, team formation, several Design Thinking workshops, elevator pitches, a teamwork workshop, a communication workshop, and a final conference with presentations. Teamwork, communication, and conflict resolution were central learning goals, supported by structured reflection tasks and team-building exercises.

All anonymised course materials are publicly available online: \url{https://figshare.com/s/d49f4dcf1c47fe82bad1}. We encourage other educators to share feedback, best practices, and ideas.

\subsection{Data Collection}
Data were collected from individual reflective assignments at the end of three courses in the summer term, using the university’s online survey tool. Courses included two research seminars ($n=61$) and one practical lab ($n=54$), with both undergraduate and graduate students. This combination allows evaluation of reflections on common pedagogical elements that are non-traditional for software engineering~\cite{meireles2024}.

Students were asked to briefly describe their experiences with the course: \emph{Please briefly describe your experiences with this course (e.g., lab, seminar). You are welcome to share anything you like as long as you are reflecting on the course itself. This can be highly individual, e.g., your perspectives on SE, the given and emergent topics, the process and implementation, teamwork, and/or your learnings. There are no right or wrong answers. It is truly about your feelings and experiences! (Not graded, Minimum 15 words)}

In addition, team reflections from the lab summarised weekly project learnings, with a particular focus on interpersonal insights. These reflections were mandatory and ungraded. They were not analysed for this report, but are mentioned for completeness.

The participation was voluntary. Students were informed about the scientific purpose of the reflections, including the plan for publication, and were asked for consent on two levels: (1) whether their team reflection could be included in the research; with the rule that if any team member declined, the whole team’s reflection was excluded from analysis, and (2) whether their individual reflection could be used for scientific work. Participation was fully voluntary, and students could opt out at any time without any consequences for their course performance. 

\subsection{Participants}
The analysed dataset includes 60 individual reflections from students across Computer Science, Business Informatics, and Cognitive Science. 
Of the 60 participants, 21 identified as women, 36 as men, and one as non-binary. Most were undergraduates: 39 in Computer Science, eight in Business Informatics, and two in Cognitive Science. Graduate students included two in Computer Science and seven in Business Informatics.

\subsection{Data Analysis}
We analysed the data using reflexive thematic analysis~\cite{braun2022}, which is well-suited for exploratory work. This approach emphasises the researchers’ active role in interpreting qualitative data and acknowledges that themes are generated rather than discovered. It is particularly appropriate for an experience report, as the first author was both course instructor and researcher, and the analysis explicitly considered how teaching experience and relationships with students might shape interpretations.
Our aim was not to quantify responses, but to capture the variety and depth of student reflections. Following Braun and Clarke’s approach~\cite{braun2022}, we deliberately avoided counting occurrences.

The first author conducted initial coding of all 60 quotes inductively, without predefined categories. Both authors discussed the initial codes. The second author independently recoded all quotes, adding more fine-grained new codes, which the first author then reviewed; both authors then agreed on the final codebook.

%steps of analysis
The analysis progressed over the standard phases~\cite{braun2022}. The analysis followed the standard phases: familiarisation with the material, initial coding (descriptive and interpretive), grouping codes into candidate themes linked to well-being, mental health, and sustainability, reviewing themes against student quotes to remain close to participants’ voices, and finalising and naming themes collaboratively. The final codebook is publicly available online \url{https://figshare.com/s/d49f4dcf1c47fe82bad1}.

\subsection{Positionality Statement}
Both authors identify as women of European origin. The first author has an interdisciplinary background that spans humanities and gender studies, as well as software engineering education. 
%\bptodo{Guter Start - darf da noch mehr rein?}

The second author has a background in sustainability research and software engineering with additional qualifications as a yoga teacher, breathwork instructor, and embodied mindfulness coach.

\section{Findings} 
\begin{figure*}[t]
	\centering
	\includegraphics[width=1\linewidth]{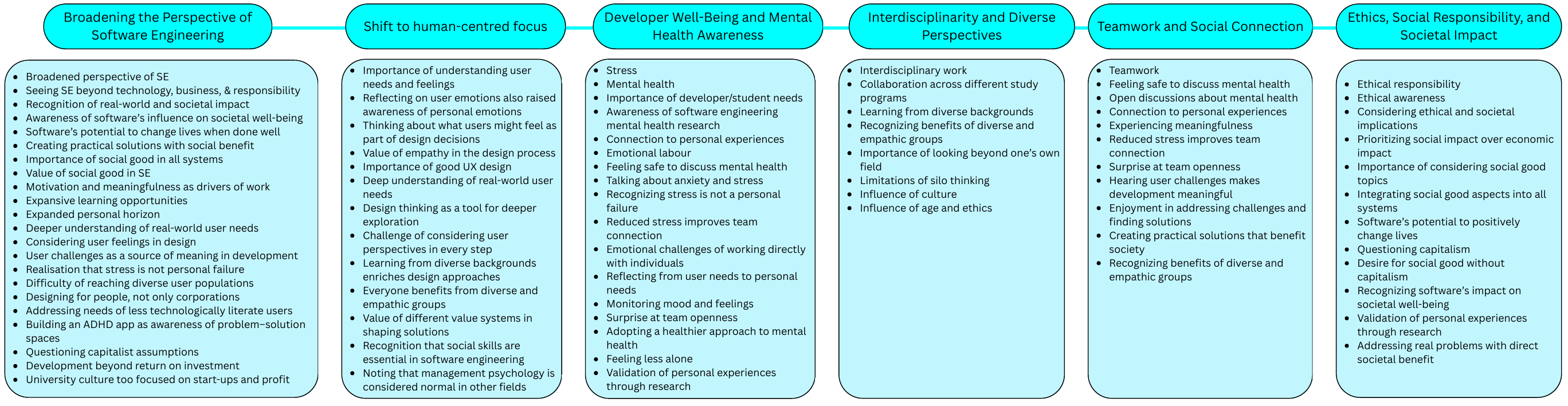}
	\caption{Overview of six themes and their corresponding codes emerged from students’ quotes.}
	\label{fig:themes}
\end{figure*}

At the end of the summer term 2025, students reflected on their experiences in the courses. We first present the findings by theme, followed by the instructor’s perspective.

\subsection{Student Voices} 
Our analysis revealed six interconnected themes that illustrate students’ learning journey. 
As shown in \cref{fig:themes}, these themes follow a narrative: students first broadened their understanding of software engineering beyond technical concerns, which naturally emerged from a human-centred focus. This shift allowed space for reflection on their own well-being as developers, while simultaneously opening them to interdisciplinary perspectives. This broadened view strengthened their appreciation for teamwork and social connection, ultimately leading to a deeper sense of ethics and social responsibility in software engineering.

\subsubsection{Broadening the Perspective of Software Engineering}
%codes: \emph{broadened SE perspective, SE more than tech/business, responsibility, real-world/societal impact, and motivation/meaningfulness, (NEW:) intention matters, widened my horizon, software impacts societal well-being, create something practical that benefits society, beneficial to consider social good aspects in all systems, thinking about what users might feel, experience of difficulty to reach the actual user population, expansive learning opportunity, deep understanding of real world user needs, realization stress is not personal failure, hearing user challenges made development feel meaningful, software can change lives when done correctly, social good, development not for ROI, not just for corporations but for people, for people who are not as technologically literate, questioning capitalism, university otherwise more concerned with start-ups and profit, Building an ADHD app, more awareness of problem-/solution-space.}
%
Students consistently described how the course fundamentally shifted their understanding of software engineering from a purely technical discipline to one embedded in broader societal contexts. This theme serves as the foundation for all subsequent experiences, as students move beyond viewing software engineering as merely coding to recognising its potential for meaningful societal impact.

Many students initially held restrictive views of their discipline. They described having \interviewquote{thought  software engineering was mostly about code and testing the code}{ID22} and feeling reduced to \interviewquote{a coding machine in all those pure technical seminars and projects.}{ID57} 
This narrow perspective might create a sense of disconnection from the broader purpose of their work as developers.

Exposure to research on well-being and sustainability revealed the potential of software engineering. Students began to see technology as a \emph{powerful} tool for addressing societal challenges: \interviewquote{I realized how powerful technology can be when applied to real societal challenges like sustainability, well-being, or accessibility.}{ID01}
Another student emphasised this expanded scope: \interviewquote{It has widened my horizon by showing me that software engineering is a powerful tool and can be used for anything, also social good topics that normally don’t get enough attention in the app/software development world.}{ID55}

This broadened understanding also challenged students’ assumptions about the commercial focus of software development, revealing alternatives to profit-driven approaches: \interviewquote{My biggest learning in this course is that software does not have to be this technical construct without feelings, used to gain money or to solve business problems.}{ID02}
This was especially due to the nature of the project topics they came up with themselves because of their interest: \interviewquote{Building an ADHD-friendly app showed me that software engineering not necessarily need to have business or profit in mind.}{ID05}

These insights led to students recognising software engineering’s potential for changing actual lives: \interviewquote{It’s not just about writing code or building apps, it’s about how what we create affects real people and communities.}{ID59}
%One student argued:
%\interviewquote{I learned that software engineering goes beyond typica[l] technical topics and rather could change actual lives when done correctly.} {ID24}

The transformation extended beyond understanding \emph{what} software engineering is to recognising \emph{how} it should be practised. Students realised that effective software engineering requires collaborative approaches, moving away from stereotypical perceptions such as the lone programmer: \interviewquote{It made me realize the depth of SE and how much it is a group project with needed social skills rather than the stereotypical idea of it.}{ID17} 
%This insight prepared them for the human-centred approach that characterises the next theme.

\begin{tcolorbox}[
  colback=gray!10,    % light grey background for the box
  colframe=gray!60,   % border color
  colbacktitle=gray!60, % darker grey background for the title
  coltitle=white,     % title text color
  title=Interpretation, % the title text
  fonttitle=\bfseries % bold title
]
Integrating well-being and sustainability into software engineering courses can \emph{inspire} students to view software as supporting societal and individual well-being, beyond technical or economic goals.\end{tcolorbox}

\subsubsection{Shift to human-centred focus}
%The codes for this theme were: \emph{importance of user needs/feelings, learning from diverse backgrounds, and design thinking, (NEW:) thinking about what users might feel, reflecting on user feelings opens eyes to own feelings, importance of good UX design, deep understanding of real world user needs, design thinking helped go deeper, difficult to consider user perspective in all steps, everyone benefits from a diverse and empathic group, value systems, empathy, normal to have management psychology in other fields, social skills needed for SE}.
%
Building on their broadened view of software engineering, students developed a deeper appreciation for human-centred design, shifting from prioritising technical functionality to recognising users’ complex needs, emotions, and lived experiences.

The shift manifested most clearly in students’ design approaches. 
Rather than beginning with technical solutions, they learned to centre user needs in their research and development process: \interviewquote{For me personally, it was a completely new perspective to step away from focusing primarily on functionality and instead place the user at the centre.}{ID10}

This user-centric approach required students to challenge their initial assumptions and remain open to discovering unexpected user needs as they learned to \interviewquote{pick up on the user’s needs and let them guide you through the design. you need to accept that your initial ideas may not be the solution, or there may not be a problem for your solution.}{ID48}

This process began with ethical considerations, and user needs identification, as one student noted: \interviewquote{it begins with ethical questions, finding user needs and solving concrete problems and all of this in a responsible way.}{ID05}

The human-centred focus also sharpened students’ critical evaluation of existing technology. They developed greater awareness of design intentions behind software interfaces: \interviewquote{I was able to see how much thought goes into the work of software designers and developers, which has made me view apps and websites more critically in general. Behind every function and every design, there should be an intentional thought—a thought that considers a diverse group of users.}{ID25}

Working with diverse user populations expanded students’ social awareness. For instance, one student developed \interviewquote{an awareness of the problems faced by migrant and seasonal workers, because I honestly had not realised how difficult it is to reach people in this field.}{ID53}

Engaging with different lived experiences encouraged students to consider not only user emotions but also their as developers.

This outward focus on user feelings created space for internal reflection, helping students to recognise the emotional aspects of their own development work: \interviewquote{When creating that app, I realized that many detailed elements need to be considered, and I feel the need to study more aspects of how the mood and feelings of people influence the design, but also the team itself when working on the app.}{ID08}

\begin{tcolorbox}[
  colback=gray!10,    % light grey background for the box
  colframe=gray!60,   % border color
  colbacktitle=gray!60, % darker grey background for the title
  coltitle=white,     % title text color
  title=Interpretation, % the title text
  fonttitle=\bfseries % bold title
]
Exploring emotions beyond rational thinking is integral to human-centred design, helping students create more empathetic and targeted solutions.
\end{tcolorbox}

\subsubsection{Developer Well-Being and Mental Health Awareness} 
%The codes for this theme were: stress, importance of developer/student needs, open to discuss mental health, awareness of SE mental health research, connection to own experiences, and emotional labour.
%(NEW:) mental health, feeling safe to discuss here, talk about anxiety and stress, realizing stress is not personal failure, less stress improved team connection, working directly with individuals was emotionally challenging, reflection from user needs to own needs, study mood and feelings, surprise about openness in team, healthier approach to subject, feeling less alone, validation of own experience by seeing research about it
%
As students focused on user well-being, this concern turned inward, leading to reflection on their own mental health and stress levels as developers. 
This theme now illustrates how considering well-being topics created a mirror effect, encouraging students to reflect on their own emotional needs and work practices.

Working on mental health applications provided particularly powerful learning experiences. Students connected their project work with personal insights: \interviewquote{Working on mental health apps made me realize how much pressure I put on myself as a student when developing some projects at university. Focusing on user well-being helped me reflect on my own stress levels and find somehow a healthier approach to the subject.}{ID11}

The integration of well-being concepts created opportunities for emotional awareness and expression. Students appreciated being able to discuss mental health openly: \interviewquote{The course has opened up my eyes to not just thinking about what users might feel, but also acknowledging what I feel as the person who then implements it and being able to transport this.}{ID33} 

This emotional literacy extended beyond individual reflection to team dynamics.
Students learned to recognise the importance of sustainable work practices. Rather than rushing through development tasks, they began valuing reflection: \interviewquote{At beginning I wanted to do everything at a same time or very fast, but I realised sometimes slowing down the process and reflecting is equally important as finishing the task.}{ID22}

The course aimed to create a \emph{safe space} for students to acknowledge and discuss stress. 
This supportive environment, with an emphasis on well-being, provided validation and a sense of belonging, particularly for students from diverse backgrounds and challenges: \interviewquote{As someone who lives with a chronic illness, it was really valuable to finally be able to talk about such things at university with others. I had often felt that no one really cared as long as everything (technical) in the project was running smoothly, and here it felt it was actually one of the main focuses.}{ID62}

Many students realised that their struggles were not isolated personal failings but common challenges in the software development community. One student felt validated by research findings: \interviewquote{Finding research on developer burnout and anxiety in the SE seminar made me feel less alone. I realized that the stress I experience isn’t just my personal failing.}{ID20}

This normalisation of mental health challenges continued as students engaged with the topic: 
\interviewquote{Reading the studies about burnout among developers was both shocking and, in a way, reassuring, because it made me feel less strange about my own experiences.}{ID46} 
The surprise many students expressed upon discovering this research highlighted gaps in typical software engineering education: 
\interviewquote{I was very surprised to see that there’s really research on how different certain environments and styles affect developer well-being, and that you not only experience this as student and feeling weird about it.}{ID27}

Beyond personal reflection, students also realised the emotional demands of engaging directly with vulnerable populations, as this created both challenges and meaning: 
\interviewquote{We made a lot of interviews, so at some point working directly with the individuals was emotionally challenging. Hearing their stories made the development process feel heavier at this point but also more meaningful than any other (purely) technical project I worked on before.}{ID37}

%\interpretation{}{Students felt better supported and reflected on their own stress }
\begin{tcolorbox}[
  colback=gray!10,    % light grey background for the box
  colframe=gray!60,   % border color
  colbacktitle=gray!60, % darker grey background for the title
  coltitle=white,     % title text color
  title=Interpretation, % the title text
  fonttitle=\bfseries % bold title
]
Students felt better supported when well-being topics were explicitly addressed in software engineering courses, allowing reflection on their own stress and mental health.
\end{tcolorbox}

\subsubsection{Interdisciplinarity and Diverse Perspectives}
%The codes for this theme were: interdisciplinary work, different study programs, and learning/seeing diverse backgrounds. (NEW:) everyone benefits from a diverse and empathic group, immense importance of looking beyond own field, silo thinking is insufficient, culture, age and ethics has influence
%
Students’ growing awareness of human needs and their own well-being fostered appreciation for interdisciplinary collaboration and diverse perspectives. They recognised the limits of purely technical approaches and embraced insights from multiple fields.

Students discovered that meaningful software solutions require input from various fields beyond computer science: \interviewquote{Technical solutions can miss the point when they are not informed by diverse fields. Social sciences, art, policy, and lived experience must shape technical choices. Changing the viewpoint is an important step for every development.}{ID06} 
This recognition challenged the pure technical understanding: \interviewquote{I came to understand why isolated knowledge and thinking are insufficient, and that the more diverse people you have in a room, and the more mutual understanding there is, the more everyone benefits.} {ID49} 

Collaboration with students from other disciplines proved particularly valuable since \interviewquote{the more diverse people you have in a room, and the more mutual understanding there is, the more everyone benefits.}{ID49}  
Students appreciated learning from teammates with different academic backgrounds, noting: \interviewquote{It feels incredibly important to look beyond one’s own perspective (as one teammate from cognitive science put it), because it completely changes how you see things and allows you to learn a lot. Afterwards, you really feel that you know and understand more than before.}{ID14}

Students with interdisciplinary backgrounds often served as bridges, helping technically focused teammates appreciate the value of diverse approaches. One business informatics student reflected: \interviewquote{I’m used to have this kind of other approaches also in business as there are whole subjects only focusing on management psychology or training.}{ID23}  

These experiences enriched the learning for all team members. 
Students with prior interdisciplinary experience felt validated for their beliefs about collaborative approaches: \interviewquote{I have worked or been interested in UX and design processes generally before the course, [...] because the importance of good design is often overlooked by computer scientists.}{ID15}

\begin{tcolorbox}[
  colback=gray!10,    % light grey background for the box
  colframe=gray!60,   % border color
  colbacktitle=gray!60, % darker grey background for the title
  coltitle=white,     % title text color
  title=Interpretation, % the title text
  fonttitle=\bfseries % bold title
]
Interdisciplinary perspectives raised students’ awareness of cultural backgrounds, mental health, and other needs, creating mutual learning benefits for all disciplines.
\end{tcolorbox}

\subsubsection{Teamwork and Social Connection}
%The codes for this theme were: teamwork, open to discuss mental health, connection to own experiences, meaningfulness.(NEW:)less stress improved team connection, surprise about openness in team, feeling safe to discuss here, hearing user challenges made development feel meaningful, it’s fun to address challenges and find solutions, create something practical that benefits society, everyone benefits from a diverse and empathic group
%
The emphasis on diverse perspectives enhanced students’ appreciation for collaborative work within development teams. This theme illustrates how students discovered that effective software engineering requires empathy, trust, and open communication among team members.

Students challenged stereotypical notions of software engineering as a solitary endeavour. They came to understand that successful projects depend on collaborative skills and social awareness: \interviewquote{It made me realize the depth of SE and how much it is a group project with needed social skills rather than the stereotypical idea of it.}{ID17} 

The course’s reduced emphasis on technical pressure created space for genuine team connection. Students found that \interviewquote{Since the project and the team was not that stressed about the whole technical setup and pressure of getting it working, it actually improved our ability to connect as a team and my own work-life balance.}{ID29} 

This supportive environment fostered deeper connection and mutual support. 
Open discussions about mental health reduced stigma and strengthened team bonds. Sharing experiences across diverse backgrounds promoted understanding. One cognitive science student noted how 
\interviewquote{seeing working on apps with the other guys with their solely software engineering background who opened up then, made my team more comfortable sharing our own mental health struggles.}{ID32}

The integration of well-being concepts helped students recognise that supporting team members’ emotional needs contributes to project success. They learned that acknowledging vulnerability and providing mutual support creates stronger, more effective teams: \interviewquote{it is okay to feel overwhelmed and down and just talk to the others.}{ID57}

Students developed greater confidence in their collaborative abilities. They learned to overcome intimidation: \interviewquote{I learned I don’t need to be intimidated [...] We’re all on the same team, and if we trust and rely on each other, we can eventually finish the project and deliver results.}{ID63} This confidence-building was particularly important for students who felt technically less prepared.

\begin{tcolorbox}[
  colback=gray!10,    % light grey background for the box
  colframe=gray!60,   % border color
  colbacktitle=gray!60, % darker grey background for the title
  coltitle=white,     % title text color
  title=Interpretation, % the title text
  fonttitle=\bfseries % bold title
]
Empathic communication is crucial in interdisciplinary software engineering teams, and integrating it helped students reflect on team interactions and well-being.
\end{tcolorbox}

\subsubsection{Ethics, Social Responsibility, and Societal Impact}
%The codes for this theme are:  ethical responsibility, ethical awareness, ethical and societal implications, social impact over economic impact, important to consider social good topics, beneficial to consider social good aspects in all systems, software can change lives when done correctly, questioning capitalism, wish for social good without capitalism, software impacts societal wellbeing, validation of own experience by seeing research about it, addressing real problems, useful for the masses directly
%
The students’ learning journey is building up to an increased sense of ethics and social responsibility in software development. This final theme represents how students’ insights into a mature understanding of software engineering’s societal impact.

Students developed a sophisticated understanding of their responsibilities. They recognised software engineering \interviewquote{as a discipline with significant ethical, social, and environmental responsibilities. It highlighted how code and systems can directly impact communities, shape policy, and influence equity.}{ID56} 

This awareness transformed their professional identity and future aspirations.
Students gained an appreciation for the consequences of technological decisions. 
They developed awareness that \interviewquote{software can have social, ethical, and societal consequences, and that sustainable, inclusive, and responsible development is an integral part of good practice. In particular, the realisation that technology can also have unintended side-effects sharpened my awareness of the need for ethical reflection in the development process.}{ID50}

Students embraced the idea that \interviewquote{ethical, inclusive, and impactful design should be a core part of the development process and not an afterthought.}{ID01} 

Although ethics was not explicitly addressed in the courses, students proactively engaged with ethical considerations. They began to see themselves as active agents shaping technology’s societal role, recognising their responsibility to consider broader implications and to promote inclusive, ethical software development. 

\begin{tcolorbox}[
  colback=gray!10,    % light grey background for the box
  colframe=gray!60,   % border color
  colbacktitle=gray!60, % darker grey background for the title
  coltitle=white,     % title text color
  title=Interpretation, % the title text
  fonttitle=\bfseries % bold title
]
Integrating well-being concepts naturally fostered ethical reasoning and social responsibility in software engineering students.
\end{tcolorbox}

%\begin{itemize}
%    \item ways of working and communicating in different disciplines (examples: reflective writing, speaking about emotions, allowing oneself to be vulnerable) 
%    \item perception of software engineering from students outside SE (positive perception of having this topic included in SE since the stereotypes about software engineers still exist, but also then learning from SE team mates that this is not their usual way of working) 
 %   \item $\rightarrow$ reflected across themes  but maybe should be a own theme or at some point be reflected in discussion? with reflexive thematic analysis we could argue that it is its own theme
%\end{itemize}

\subsection{Instructor Perspective}

\interviewquote{From the very beginning, I deliberately gave students a great deal of freedom in both topic choice and project design. This was a deliberate decision, as in most of their courses, students are used to being given very specific tasks: in seminars, often with predefined research questions, and in practical courses, a list of requirements to implement. 
By contrast, in our courses, they first had to ask themselves what genuinely interested \textbf{them} in software engineering. They also had to negotiate these interests with their teammates to arrive at a shared interest. While this freedom initially overwhelmed some, the process of reflection and conversation early on seemed to strengthen their motivation. Once they had found a topic, they were more emotionally invested and committed to producing meaningful results, not only to satisfy the instructor or to receive a grade, but because they themselves really wanted to see an outcome. \\
The creative freedom was taken up very positively. Students appreciated the wide variety of subjects that became possible, ranging from technical explorations to strongly human-centred and socially engaged topics. In this sense, freedom helped to surface interests that would otherwise remain hidden. In individual meetings, I felt that this also allowed contributions from students who were less technically proficient or from underrepresented groups. Several of these students appeared to feel more included, more confident, and more willing to share ideas and continue engaging with computer science. As a result, several students even approached me afterwards to ask about pursuing a thesis topic in these areas, as their interest had been sparked during the course.\\
Another element that shaped the learning atmosphere was my decision to bring in current research outcomes as well as my own experiences from software engineering conferences. These covered interventions related to topics such as gender or well-being. I shared some of my own experiences with stress, deadlines, and collaboration in research projects. I discussed how communication changes under pressure, how misunderstandings can easily arise in online exchanges, and how humour sometimes helps to mitigate difficult situations. I shared a short excerpt from a conversation with a co-author during a tight deadline to show how such moments can be navigated. My impression was that these anecdotes made me more approachable and encouraged students to open up about their own challenges.  \newline
I believe that both the openness of topic choice and my personal sharing had a positive effect on the learning atmosphere. Students described the space as a safe environment where they could explore new themes with curiosity, without fear of judgment, and I also saw this implemented during the discussion and the feedback the students gave each other as they were so engaged. %For me as an instructor, 
The most rewarding outcome was to see students develop stronger reflection skills,  about themselves, their teamwork, and that they indeed have their  role in shaping technology in society.}{Instructor}

\begin{tcolorbox}[
  colback=gray!10,    % light grey background for the box
  colframe=gray!60,   % border color
  colbacktitle=gray!60, % darker grey background for the title
  coltitle=white,     % title text color
  title=Interpretation, % the title text
  fonttitle=\bfseries % bold title
]
The instructor shares their intention to embed important skills that contribute to emotional intelligence, especially self-awareness, empathy, and effective communication. Integrating recent research on these topics within software engineering shows their relevance beyond education.
The instructor also notices that creative freedom made the classes more inclusive.
Finally, personal stories bring the concepts to life and make the instructor approachable.
\end{tcolorbox}

\section{Discussion}
We contextualise our results within prior work, reflect on lessons learned, and outline recommendations and future research.

\subsection{Results in Light of Related Work}
Our experiences show that integrating mental health, well-being, and sustainability into software engineering education can broaden students’ understanding of the field and foster reflection on both technical and personal practices. Our findings extend the discussion on interdisciplinary perspectives in software development~\cite{hyrynsalmi2025}. 

The students’ reflections echo earlier work highlighting the importance of addressing mental health in engineering education~\cite{asghar2024}. Students’ surprise at discovering research on developer burnout, neurodiversity, and harassment mirrors findings in other disciplines, such as life sciences and engineering, where students remain unaware of available mental health resources~\cite{navarro2025,vega2023}. Similar to these students, students in software engineering courses benefit from explicit acknowledgement of mental health challenges throughout the curriculum, rather than only at critical points~\cite{navarro2025,wright2023}.

Importantly, our findings reveal a previously undocumented bi-directional effect: students used these topics and projects to reflect on their own emotions and well-being. This connection between user well-being and developer well-being has not been described in prior studies on software engineering education. Teaching human-centred design can simultaneously create space for self-reflection and normalise conversations about mental health. Since stress often serves as a \emph{gateway} to other problems~\cite{asghar2024}, embedding self-awareness in technical coursework may foster emotional resilience without requiring separate well-being interventions that are difficult to implement during courses.

There is also a clear need to support emotional resilience systematically. Students, similar to life science and biological students~\cite{navarro2025,vega2023}, may experience high stress and often develop their own coping strategies, yet would benefit from structured guidance on self-care~\cite{godfrey2011}. Despite their eagerness to go through the \emph{struggle}, this is particularly important for underrepresented groups~\cite{vorderwulbeke2025,boman2024}.

Finally, our experiences highlight that software engineering education must also prepare students for the professional challenges described as the \emph{dark side} of software engineering~\cite{rost2011dark}. Stress, lack of transparency, toxic collaboration patterns, and neglect of ethical considerations can lead to misconduct and project failure~\cite{hatton2015,raman2020stress}. By addressing mental health, resilience, and social responsibility during students’ education, we can equip them with strategies to recognise and resist these harmful dynamics. This preparation not only supports individual well-being but also strengthens the long-term sustainability and integrity of the software engineering profession~\cite{suarez2024}.

%\bptodo{Can we use this somewhere? The dark side of software engineering: evil on computing projects~\cite{rost2011dark}} \igtodo{yes, i also believe this is a perfect fit as conclusion in the discusison.}

\subsection{Trustworthiness and Limitations}
Since the students’ reflections are data in natural language collected through what can be classified as naturalistic enquiry~\cite{lincoln1985naturalistic}, we apply the trustworthiness framework by Lincoln and Gupta~\cite{lincoln1985naturalistic} to critically examine our research. They devised criteria that parallel those of the conventional paradigm. Trustworthiness is broken down into credibility, dependability, confirmability, and transferability, which are respectively equivalent to the criteria of internal validity, reliability, objectivity, and external validity or generalizability.

\paragraph{Credibility} We followed recommended strategies~\cite{lincoln1985naturalistic} of prolonged engagement (collecting written reflections over one semester), persistent observation during classes, and peer debriefing.
\paragraph{Dependability} An external audit was conducted by the second author, who joined after the courses were completed as a neutral observer to support the analysis. 
\paragraph{Confirmability} An `external audit’ by a competent external, disinterested auditor is provided via the peer-review process. Their examination of this contribution and the data results in a confirmability judgement. 
\paragraph{Transferability} We offer (not-so-`thick’) descriptive data with a narrative developed about the context so that judgments about the degree of fit or similarity may be made by others who may wish to apply all or part of the findings elsewhere. The explication of \emph{not-so-`thick’} refers to the original strategy by Lincoln and Gupta requesting the descriptive data to be `thick’, whereas the analysis of our course instances relies on a limited set of written reflections. Since our contribution is an experience report based on a pilot rather than a fully fledged empirical study, we deem it sufficient to offer initial insights and do not claim transferability.

The limitations of this report are that the data stem from a single university, our analysis is based on self-reported reflections and lacks systematic measurements, and there is a self-selection bias because the students actively chose to enrol in these (elective) courses. 
For transparency and reproducibility, the final codebook and anonymised course materials are publicly available online.\footnote{\url{https://figshare.com/s/d49f4dcf1c47fe82bad1}}

\subsection{Lessons Learned \& Recommendations}
Student reflections highlight two key points: (1) the stereotype of the purely technical software engineer persists, and (2) non-technical or \emph{soft} skills, such as ethics, sustainability, well-being, and teamwork, are not merely \emph{nice to have}; they help students see software engineering as socially responsible, ethical, and inclusive. Importantly, students naturally reflected on their own mental health while working on user-centred projects, and interdisciplinary collaboration, combined with a mental health focus, created strong, bidirectional learning effects.

\subsubsection{Broaden the Scope of Software Engineering}
From our experience, students appreciate software framed as more than code and business; they feel motivated when they see its societal and ethical impact. Rather than waiting for dedicated modules or citing overloaded curricula, we found that even small integrations into existing courses make a visible difference. Examples include green software practices, developer burnout, accessibility design, AI, testing, and current real-world trends. 

$ \textcolor{cyan}{\boldsymbol{\rightarrow}}$ Include at least one topic, paper, reading, reflection, or case study on well-being, mental health, or sustainability.

\subsubsection{Create Space for Human-centred Approaches.}
We found that giving students opportunities to focus on user needs shifts their thinking from purely technical to human-centred, and encourages them to reflect on the impact of their work. This also requires students to justify decisions based on user needs. 

$ \textcolor{cyan}{\boldsymbol{\rightarrow}}$ Incorporate at least one user activity per technical project, such as interviews, persona creation, or accessibility evaluation.

\subsubsection{Normalise Mental Health Discussions.}
We observed that many students linked course tasks to their own stress and were surprised by openly talking about stress, overwhelming, anxiety and burnout. 
We believe this is exactly what a university is supposed to be: an environment where students \emph{learn} how they approach stressful situations. The reflections will help them to better understand their feelings and actions. This signals that developer well-being matters and helps students understand their own coping strategies.

$ \textcolor{cyan}{\boldsymbol{\rightarrow}}$ Include short reflections (weekly journals, stand-ups, retrospectives) asking: \emph{How did this project affect your stress or well-being?}

\subsubsection{Normalise Vulnerability and Authentic Communication}
We discovered that sharing our personal challenges with stress and deadlines makes students more willing to engage openly and builds a safer, more authentic classroom culture. 
Discuss how stress affects decision-making, how misunderstandings arise in technical collaborations, and how experienced practitioners navigate difficult situations. This humanises the instructor and normalises discussions about challenges inherent in software development.

$ \textcolor{cyan}{\boldsymbol{\rightarrow}}$ Share and discuss brief, relevant anecdotes about professional challenges, stress, or collaboration. 

\subsubsection{Strengthen Teamwork and Social Connection}
We learned that prioritising team-building before technical work helps students form stronger connections, reduces feelings of isolation, and supports collaboration. Use informal activities (e.g., a pizza lunch) to foster trust, which is especially relevant for diverse teams that may need additional time to move through the forming and storming phases. Students need to see \emph{and} learn that connecting with teammates is as important as setting up technical infrastructure. 

$ \textcolor{cyan}{\boldsymbol{\rightarrow}}$ Begin courses with mandatory team-building exercises and contracts, and emphasise informal activities throughout the course.

\subsubsection{Integrate Interdisciplinary and Ethical Perspectives.}
We realised that exposing students to interdisciplinary insights helps them understand the societal impact of software and encourages more responsible decision-making. In an ideal scenario, you could invite one guest to provide input from a real person. However, without a major redesign, you can also include an interdisciplinary paper or approach (e.g., qualitative methods or interviews) from another field (e.g., the psychology of teamwork, accessibility design, or sustainability). This should be framed as standard engineering considerations rather than a philosophical side note. Therefore, you would then also incorporate the first two recommendations from above. 

$ \textcolor{cyan}{\boldsymbol{\rightarrow}}$ Encourage students to consider broader societal implications, by asking \emph{Who might be excluded by this design?} and \emph{What could go wrong if this system were misused?}

\subsubsection{Course Content Update}
Based on our pilot experience, future iterations of our courses should include: 

$ \textcolor{cyan}{\boldsymbol{\rightarrow}}$ Explicitly state that courses develop both technical competence and social awareness (self-reflection, empathetic design, ethical reasoning) as learning objectives. 

$ \textcolor{cyan}{\boldsymbol{\rightarrow}}$ Maintain creative freedom in topic choice, but provide early guidance and reflection discussions to reduce students’ feeling of being overwhelmed.

$ \textcolor{cyan}{\boldsymbol{\rightarrow}}$ Expand classroom discussions with structured formats for large courses (tandem discussions, small group rotations, online forums with peer feedback).

$ \textcolor{cyan}{\boldsymbol{\rightarrow}}$ Replace traditional presentations with poster sessions to encourage informal discussion and peer learning.

$ \textcolor{cyan}{\boldsymbol{\rightarrow}}$ Include practical stress management techniques (breathing exercises, mindfulness) in project management training.

$ \textcolor{cyan}{\boldsymbol{\rightarrow}}$ Integrate contemporary challenges such as AI ethics and sustainability~\cite{canavate2025}.

\subsection{Future Work}
Our experience report represents the basis for a systematic empirical investigation. We plan to investigate the following research questions using mixed-methods approaches.

%$\textcolor{cyan}{\bigstar}$  How does exposure to mental health topics in software engineering courses change students’ views of what the field can do? \bptodo{add the how}

$\textcolor{cyan}{\bigstar}$ How do well-being interventions (e.g., mindfulness, breathing exercises, reflective journaling) influence students’ stress management and learning in software projects? We will combine validated psychological instruments with qualitative reflections to assess changes in stress, coping, and teamwork.

$\textcolor{cyan}{\bigstar}$ How do students’ perceptions of stress and resilience vary across diverse cultural backgrounds, disciplines, and personal experiences? We will investigate how differences in prior experiences, attitudes, and societal expectations shape engagement with well-being-focused activities.

$\textcolor{cyan}{\bigstar}$ How do students navigate interactions with peers who are less interested or sceptical about well-being and social aspects of software engineering? We will explore strategies for maintaining engagement and fostering discussions, particularly in team settings with varying motivation.

$\textcolor{cyan}{\bigstar}$ How do remote or hybrid settings~\cite{suarez2024} affect students’ resilience, social connection, and ability to reflect on well-being? We will analyse how different modes of study impact personal coping strategies, collaboration, and peer support networks.

%$\textcolor{cyan}{\bigstar}$ Do students become more aware of mental health resources and develop better coping strategies after working on well-being projects? We will track students’ knowledge of campus resources, use of support services, and development of personal coping strategies through surveys and interviews.

%$\textcolor{cyan}{\bigstar}$ Do classroom discussions about mental health in technical contexts make students more willing to seek help and discuss personal problems? \bptodo{add the how}

$\textcolor{cyan}{\bigstar}$ How do the changes from the pilot to the above proposed version reflect in the students’ reflections? We will conduct a comparative thematic analysis of team reflections from the pilot and updated course versions to identify improvements and persistent challenges.

\section{Conclusions}
This research addresses the under-explored area of mental health, well-being, and sustainability in software engineering education. By documenting our teaching approach and exploring its effects, we provide guidance for educators seeking to foster more supportive learning environments that expand students’ professional horizons. 

Students’ learning journey progressed from narrow technical conceptions through human-centred awareness, personal well-being reflection, interdisciplinary appreciation, to seeing software engineering as a socially responsible discipline. This demonstrates that even small interventions can have a positive initial impact on both personal and professional identity.

\bibliographystyle{ACM-Reference-Format}
\bibliography{references}

\end{document}